\documentclass{aa}  
\newcommand{\vc}[1]{{\boldsymbol #1}}
\usepackage{graphicx, bm}
\usepackage{txfonts}

\begin{document} 

\title{Effect of nucleation on icy pebble growth \\in protoplanetary discs}

   \author{Katrin Ros\inst{1}
          \and
          Anders Johansen\inst{1}
          \and
          Ilona Riipinen\inst{2,3,4}
           \and
           Daniel Schlesinger\inst{2,3}
          }

   \institute{Lund Observatory, Department of Astronomy and Theoretical Physics,
              Lund University, Box 43, 221 00 Lund, Sweden\\
              \email{katrin.ros@astro.lu.se}
             \and
             Department of Environmental Science and Analytical Chemistry, Stockholm University, SE-10691 Stockholm, Sweden
             \and
             Bolin Centre for Climate Research, SE-10691 Stockholm, Sweden
             \and
             Aerosol Physics, Faculty of Science, Tampere University of Technology, Tampere, Finland}

   \date{Received 27 September 2018; Accepted 12 July 2019}

 \abstract
{Solid particles in protoplanetary discs can grow by direct vapour deposition outside of ice lines. The presence of microscopic silicate particles may nevertheless hinder growth into large pebbles, since the available vapour is deposited predominantly on the small grains that dominate the total surface area. Experiments on heterogeneous ice nucleation, performed to understand ice clouds in the Martian atmosphere, show that the formation of a new ice layer on a silicate surface requires a substantially higher water vapour pressure than the deposition of water vapour on an existing ice surface. In this paper, we investigate how the difference in partial vapour pressure needed for deposition of vapour on water ice versus heterogeneous ice nucleation on silicate grains influences particle growth close to the water ice line. We developed and tested a dynamical 1D deposition and sublimation model, where we include radial drift, sedimentation, and diffusion in a turbulent protoplanetary disc. We find that vapour is deposited predominantly on already ice-covered particles, since the vapour pressure exterior of the ice line is too low for heterogeneous nucleation on bare silicate grains. Icy particles can thus grow to centimetre-sized pebbles in a narrow region around the ice line, whereas silicate particles stay dust-sized and diffuse out over the disc. The inhibition of heterogeneous ice nucleation results in a preferential region for growth into planetesimals close to the ice line where we find large icy pebbles. The suppression of heterogeneous ice nucleation on silicate grains may also be the mechanism behind some of the observed dark rings around ice lines in protoplanetary discs, as the presence of large ice pebbles outside ice lines leads to a decrease in the opacity there.}


   \maketitle
%

\section{Introduction}

Our understanding of how planets form in protoplanetary discs proves to be insufficient already at the stage of growth from micrometre-sized dust grains to centimetre- to decimetre-sized pebbles. Small dust particles grow efficiently by coagulation, but at sizes above millimetres the sticking of particles in collisions is replaced by bouncing and fragmentation \citep{blumwurm2008, guttleretal2010}. Furthermore, particles drift radially inwards as they lose angular momentum due to interaction with the pressure-supported gas orbiting with sub-Keplerian velocities \citep{weidenschilling1977}. This so-called radial drift barrier affects particles in the centimetre to metre range the most, causing them to rapidly drift inwards and sublimate in the vicinity of the central star \citep{braueretal2008, birnstieletal2010}. Despite these hurdles in their formation path, growth to pebble-sizes has been inferred in several star-forming regions by observations of the spectral energy distributions of protoplanetary discs \citep{wilneretal2005, riccietal2010, testietal2014}. From a theoretical perspective, pebbles are also a necessary ingredient in the further growth towards planets through the streaming instability, in which particles clump together and subsequently collapse gravitationally to form planetesimals with a characteristic size of 100 km \citep{youdingoodman2005, johansenetal2007, johansenetal2015, simonetal2016, simonetal2017, schaferetal2017, abodetal2018}, and for continued growth to planets by pebble accretion \citep{johansenlacerda2010, ormelklahr2010, lambrechtsjohansen2012, idaetal2016, johansenlambrechts2017}. 

Ice lines, locations in the disc where a certain volatile species transitions from vapour to solid ice, have the potential to be favourable locations for the growth of pebbles. Several factors contribute to this. Firstly, outside the ice lines of major volatile species the solid density increases significantly; just outside of the water ice line located at 1--3 au, depending on the evolutionary stage of the protoplanetary disc \citep{martinlivio2012, bitschetal2015}, the solid density approximately doubles, improving the conditions for fast particle growth \citep{lodders2003}. Secondly, particles covered in water ice might survive, and even grow, in collisions with ten times higher velocities than bare silicate particles \citep{wadaetal2009, gundlachblum2015}. However, recent experiments have shown that the surface energy, which is responsible for the `stickiness' of water ice, might have been overestimated for low temperatures \citep{gundlachetal2018,musiolikwurm2019}. Particles covered in $\rm CO_2$ ice have sticking properties similar to silicates and hence stall their growth at an equivalent bouncing barrier \citep{musioliketal2016A,musioliketal2016B}. In the case of water ice facilitating growth this gives rise to an optimal point with favourable collisional growth in the region between the water ice line and the $\rm CO_2$ ice line, situated at about 10 au (where $T\approx70\,\rm K$) from the central star \citep{mousisetal2012}, as demonstrated by \citet{okuzumitazaki2019}. Finally, particles outside the ice line grow as vapour is deposited\footnote{In the astronomical literature, `condensation' is often used to describe the phase transition from vapour to solid, however in this paper we use the more technically correct term `deposition' for this process.} on them, and sublimation of icy particles provides material in the form of vapour. In this paper we focus on these two latter processes: deposition of water vapour and sublimation of ice at the water ice line. 

Early work on particle growth at the ice line includes \citet{stevensonlunine1988}, who proposed a `cold-finger' effect where the vapour reservoir in the inner disc diffuses outwards across the water ice line and is deposited on solids outside the ice line. They found a large enhancement of solids beyond the ice line, but ignored dynamical processes in the disc such as the inwards transport of solids, and also assumed homogeneous nucleation of snowflakes followed by fast coagulation into large planetesimals. \citet{cuzzizahnle2004} found an enhancement of solid density just outside of the ice line of at least an order of magnitude, by considering rapid inwards drift of solids followed by a phase of vapour diffusing back out across the ice line and being deposited on immobile sink particles. In \citet{rosjohansen2013} we considered a more complex dynamical model, where vapour and small ice particles are coupled to the gas and thus are subject to turbulent diffusion, and larger particles sediment towards the midplane and drift radially inwards. By adopting a Monte Carlo model for sublimation and deposition, icy particles were found to grow to decimetre-sized pebbles in only a few thousand years as particles drift inwards and sublimate, and part of the resulting vapour diffuses outwards and is deposited on icy particles outside of the ice line. However, in \citet{rosjohansen2013} we neglected the important effect of the silicate ice nuclei that are released upon sublimation. If water vapour can be deposited equally well on ice and silicate surface, then the inclusion of these ice nuclei should greatly diminish the efficiency of growth by deposition. Recent work that included the release of small silicate dust grains upon sublimation of icy pebbles has shown a pile-up of particles around the ice line. \citet{schoonenbergormel2017} assumed a steady inflow of pebbles into their simulation region around the water ice line, and found a pile-up of solids outside the ice line of at least a factor of three, and \citet{idaguillot2016} found a significant pile-up of small particles inside the ice line due to the release of dust particles upon sublimation of the icy mantles. However, neither of these papers accounted for the efficiency with which the first ice layer forms on the silicate cores, thereby ignoring an important factor that can decrease growth significantly.
 
In this paper we include a distinction between the heterogeneous or depositional ice nucleation of water ice on a silicate surface and the deposition of water vapour on an existing water ice surface. Heterogeneous nucleation, which is the formation of a new ice phase from metastable (supersaturated) vapour on an existing substrate, can be described by classical nucleation theory \citep[CNT,][]{vehkamaki2006}. Ice clusters that form on a silicate substrate readily sublimate due to their high level of curvature, but can become stable once they grow by stochastic collisions past a critical size. The critical cluster size and hence the nucleation rate depends strongly on vapour supersaturation. Classical nucleation theory therefore predicts that a significant supersaturation is required in order for the clusters to merge and form the first full monolayer of ice -- as compared with simple deposition where net ice growth is achieved once the saturation ratio exceeds unity. In homogeneous nucleation where new ice particles form without the presence of a substrate, even higher supersaturations are required. 

Heterogeneous ice nucleation has been investigated experimentally in the context of ice cloud formation in the Martian atmosphere. The  temperature in the upper troposphere of the Martian atmosphere lies in the range 150--170 K and is hence very similar to the temperature at the ice line of a protoplanetary disc. Iraci et al. (2010) demonstrated that the onset of heterogeneous ice nucleation requires significant saturation levels at these temperatures, up to 300\%. These experimental values are two to three times higher than those yielded by classical nucleation theory using a single contact angle value, although the CNT predictions on heterogeneous ice nucleation are expected to be highly uncertain \citep[e.g.][]{vehkamaki2006}. Iraci et al. (2010) further demonstrated that this heterogeneous nucleation barrier exists for several types of terrestrial dust, even for clays that are otherwise known to absorb water molecules well at higher temperatures. In this paper we fit the data from \citet{iracietal2010} using a distribution of contact angles \citep[see][and Appendix \ref{app:nucleation}]{wheelerbertram2012} to yield a description that is more realistic than the traditional CNT predictions with a single contact angle.

The potential inhibition of heterogeneous ice nucleation on silicate dust has important implications for the growth of icy pebbles in protoplanetary discs. We demonstrate in this paper that icy pebbles grow to centimetre sizes by deposition of water vapour diffusing outwards from the water ice line and that these pebbles coexist with a population of ice-free silicate dust particles. The need for large supersaturation has been inferred for the heterogeneous nucleation of $\rm CO_2$ ice on both silicate and water ice substrates as well \citep{glandorfetal2002} and a similar inhibition may apply to the heterogeneous nucleation of CO ice on a substrate of $\rm CO_2$ ice. We therefore speculate that some of the dark rings observed in protoplanetary discs in millimetre-wavelengths \citep{almapartnershipetal2015, andrewsetal2016, tsukagoshietal2016, isellaetal2016, ciezaetal2017, andrewsetal2018} and in scattered light \citep{vanboekeletal2017} could represent a reduction in the opacity due to the growth of icy pebbles, as proposed by \citet{zhangetal2015, zhangetal2016}, and that this growth is a direct consequence of the inhibition of heterogeneous nucleation.

The paper is organised as follows. We describe the model in Sect.\ \ref{sec:method}, including the concepts of heterogeneous ice nucleation and depositional ice growth in Sect.\ \ref{sec:nucleationcondensationandsublimation}, and present the results in Sect.\ \ref{sec:results}. In Sect.\ \ref{sec:discussion} we discuss the model assumptions and implications of our results, and we conclude with our main findings in Sect.\ \ref{sec:conclusions}. In Appendix \ref{app:diffusion} we describe how we model turbulent diffusion of vapour and particles and in Appendix \ref{app:condensation} how the particle size evolution due to deposition and sublimation is modelled. In Appendix \ref{app:nucleation} we fit the experimental data used in our simulations within the framework of classical nucleation theory. 

   
\section{Method}
\label{sec:method}

\begin{table*}
\caption{Model parameters for the different runs presented in this paper. The first two columns give the model and run. The critical saturation ratio needed for heterogeneous nucleation is given in the third column, and can be either temperature dependent, in the full temperature-dependent nucleation model where we differentiate between heterogenous nucleation and depositional growth, or $S_{\rm crit, \,sil}=1$ for the simple model where these two processes are treated equally. The initial particle sizes, turbulent $\alpha$-parameter, number of particles, and scale height for water vapour and particles are given in Columns 4--7. The number of grid cells in the radial direction is $n_{\rm r}=2\times10^3$ for all runs.}             
\label{table:1}      
\centering                                      
\begin{tabular}{l l l l  l l l}         
\hline\hline                        
Model & Run & $S_{\rm crit,\, sil}$ & $a_{\rm init}$ $(\rm cm)$  & $\alpha$ & $n_{\rm p}$ & $H_{\rm H_2O}$ \\
\hline                                   
 Full temperature-dependent nucleation model  & Fiducial & T-dep & 0.1 & $10^{-3}$ & $2\times10^6$ & $0.1\,H_{\rm g}$\\ 
 &  mm \& $\mu$m & T-dep &  0.1 and 0.0001 & $10^{-3}$ & $2\times10^5$ & $0.1\,H_{\rm g}$\\
 &  Low resolution & T-dep & 0.1 & $10^{-3}$ & $2\times10^5$ & $0.1\,H_{\rm g}$ \\ 
 & Low $\alpha$ & T-dep & 0.1 & $10^{-4}$ & $2\times10^6$ & $0.1\,H_{\rm g}$\\
  & Large $H_{\rm H_2O}$ & T-dep & 0.1 & $10^{-3}$ & $2\times10^6$ & $1\,H_{\rm g}$ \\ 
\hline 
 Simple model & mm \& $\mu$m & 1 &  0.1 and 0.0001 & $10^{-3}$ & $2\times10^5$ & $0.1\,H_{\rm g}$\\
\hline                                             
\end{tabular}
\end{table*}

We ran simulations where we included water ice, vapour, and silicate dust particles and compared the resulting particle sizes in a simple and a full temperature-dependent nucleation model. In our simple model all particles have the same flux of water ice deposition onto them - we do not differentiate between heterogeneous ice nucleation and continued growth by deposition. In the full nucleation model we do distinguish between  heterogeneous nucleation and continued depositional ice growth by implementing the experimental results of \citet{iracietal2010}, which imply that vapour is preferably deposited on already ice-covered surfaces. We want to isolate the effects of deposition and sublimation in a dynamical model and therefore ignore other potential growth effects, such as coagulation. Our aim is to understand how heterogeneous ice nucleation and depositional ice growth impacts the particle distribution in the vicinity of the ice line and whether these processes can lead to large enough particles to facilitate further growth into planetesimals, also when we include the abundance of potential ice nuclei in the form of silicates in the disc.

\subsection{Disc model}

We set up our disc model following the minimum mass solar nebula (MMSN) \citep{hayashi1981}, with temperature and gas column density profiles of
\begin{equation}
T=280\,{\rm K} \,\left( \frac{r}{\rm au}\right)^{-1/2}
\end{equation}
and
\begin{equation}
\Sigma=1000{\rm \,g\,cm^{-2}} \,\left( \frac{r}{\rm au}\right)^{-1}\,, 
\end{equation}
assuming a solid-to-gas ratio of $Z=0.01$, and a ratio of water vapour and ice relative to the hydrogen and helium gas of $Z_{\rm H_2O}=0.005$. These parameters yield a water ice line at $r_{\rm ice} \approx 2.6 \,{\rm au}$. In this disc, particles move in response to turbulent diffusion, sedimentation towards the midplane, and radial and azimuthal drift, while we ignore the negligible contribution from Brownian motion \citep{braueretal2008}. With the $\alpha$-prescription that describes the effectivity of angular momentum transfer in turbulent motion, introduced by \citet{shakurasunyaev1973}, the turbulent diffusion coefficient of gas and small particles in a turbulent disc can be expressed as
\begin{equation}
D=\alpha c_{\rm s}H\,.
\end{equation}
The sound speed $c_s$ increases with temperature as
\begin{equation}
c_{\rm s}=\sqrt{\frac{k_{\rm B}T}{\mu}}\,,
\end{equation}
where $k_{\rm B}$ is the Boltzmann constant and $\mu$ is the molecular weight of the gas. The gas scale height $H$ can be expressed as
\begin{equation}
H=\frac{c_{\rm s}}{\Omega}\approx0.033\,{\rm au}\left(\frac{r}{\rm au}\right)^{5/4}\,,
\end{equation}
with $\Omega$ being the Keplerian orbital frequency. We set the particle scale height, valid also for water vapour in our model, to $H_{\rm p}=0.1H $.
The radial drift velocity inwards due to the pressure-supported gas is 
\begin{equation}
v_{\rm d}=-2\,{\rm St} \Delta v\,,
\end{equation}
where $\Delta v\approx 50\,{\rm m\,s^{-1}}$ is the difference between the gas velocity and the Keplerian velocity, and
\begin{equation}
{\rm St}=\Omega\tau_{\rm f}=\frac{a\rho_\bullet}{H \rho_{\rm g}}
\end{equation} 
is the Stokes number or dimensionless friction time of the particle in the Epstein regime, with $a$ being the particle radius, and $\rho_\bullet$ and $\rho_{\rm g}$ being the material and gas density \citep{weidenschilling1977}. A more detailed description of the particle dynamics in our model is given in Appendix \ref{app:diffusion}.

\subsection{Simulation setup}

Our simulations are set up around the water ice line at $r_{\rm ice}\approx 2.6\,{\rm au}$, with a box ranging from $r_{\rm min}=2.0\,\rm au$ to $r_{\rm max}=4.5\,{\rm au}$.  Our model is one dimensional with $n_r=2000$ grid cells in the radial direction and $n_\theta=n_z=1$ in the azimuthal and vertical direction. Vapour, ice, and silicate particles are represented by $n_{\rm p}=2\times10^5$ or $n_{\rm p}=2\times10^6$ superparticles, with periodic boundary conditions being used for both solids and vapour. Each superparticle represents a total mass of $m_{\rm SP}$, divided into a large number of physical particles, where characteristics such as mass, internal density, and composition are the same for all solid physical particles represented by superparticle $i$, similar to the models by \citet{ormelspaans2008} and \citet{zsomdullemond2008}. The physical particles are assumed to be spherical, and at the start of a simulation are either millimetre sized, consistent with particle coagulation outcomes \citep{blumwurm2008, guttleretal2010}, or in the form of micrometre-sized dust. The particles consist of a rocky nucleus covered by an ice mantle, with the remaining water being distributed as vapour. This initial setup thus mimics a scenario in which thin ice mantles form on dust grains further out in the protoplanetary disc or in the protostellar cloud, and drift inwards \citep[e.g.][]{vandishoecketal2014}. The total mass fractions in the simulation region are $A_{\rm H_2O}=A_{\rm sil}=0.5\%$, approximately consistent with mass fractions in protoplanetary discs estimated from the solar nebula \citep{lodders2003}. We model a turbulent disc with $\alpha=10^{-3}$, and also include a run with $\alpha=10^{-4}$ for comparison; this range is motivated by observations \citep[e.g.][]{pinteetal2016}. 

In Table \ref{table:1} we list all our simulations, with the values of initial particle size, turbulent $\alpha$-value, number of particles, and scale height for water vapour and particles for each run. In our full nucleation model the difference between heterogeneous nucleation on silicate and continued growth by deposition on an ice surface is taken into account with a temperature-dependent $S_{\rm crit,\, sil}\geq1$, while $S_{\rm crit,\, ice}=1$. The difference between heterogeneous nucleation and depositional ice growth is discussed in more details in Sect.\ \ref{sec:nucleationcondensationandsublimation}. We also ran a simple model, where heterogeneous nucleation on silicate and deposition on ice are treated equally, with the critical saturation ratio required for heterogeneous nucleation and depositional growth being $S_{\rm crit,\, sil}=S_{\rm crit,\, ice}=1$. 

   \begin{figure}
    \resizebox{\hsize}{!}{\includegraphics{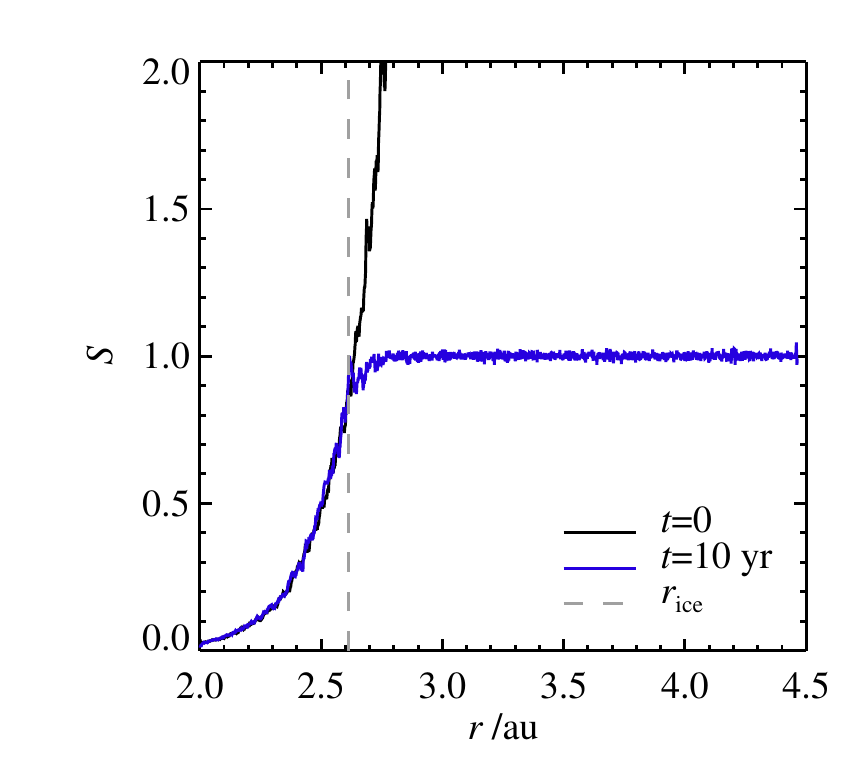}}
   \caption{Saturation ratio $S=P_{\rm v}/P_{\rm sat}$ for water vapour throughout the simulation region. Black shows the initial saturation ratio and blue shows the state after an equilibrium between vapour and ice has been reached. The ice line location at the start of the simulation, where $S$ reaches unity, is marked by the grey dashed line. Initially, all particles have a thin ice mantle, and the remaining water is distributed throughout the simulation domain as vapour, leading to a steeply rising $S$. The excess water is quickly deposited on the available solids, so that $S\approx1$ everywhere outside of the ice line, as shown for $t=10\,\rm yr$.}
              \label{fig:sr}%
    \end{figure}
 
  \begin{figure}
   \resizebox{\hsize}{!}{\includegraphics{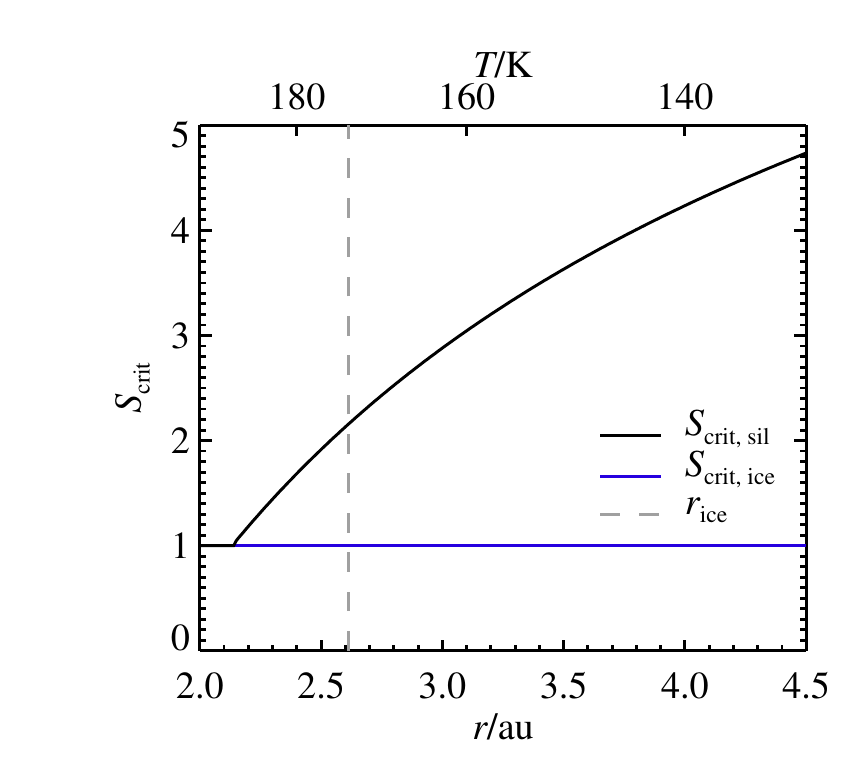}}
   \caption{Critical saturation ratio $S_{\rm crit}$ as a function of semi-major axis for the full extent of our simulations shown for bare silicate particles (heterogeneous ice nucleation) in black and water-ice-covered particles (depositional ice growth) in blue. Also shown is the ice line location, where $S=1$, as a grey dashed line. The upper horizontal axis shows corresponding disc temperatures, increasing inwards as $T/{\rm K}=280 \times (r/{\rm au})^{-1/2}$. Critical saturation ratios for silicate in the temperature range $160\leq T\leq179$ are experimental data from \citet{iracietal2010}.}
              \label{fig:srt}%
    \end{figure}

\begin{figure}
{\includegraphics[width=8.8cm]{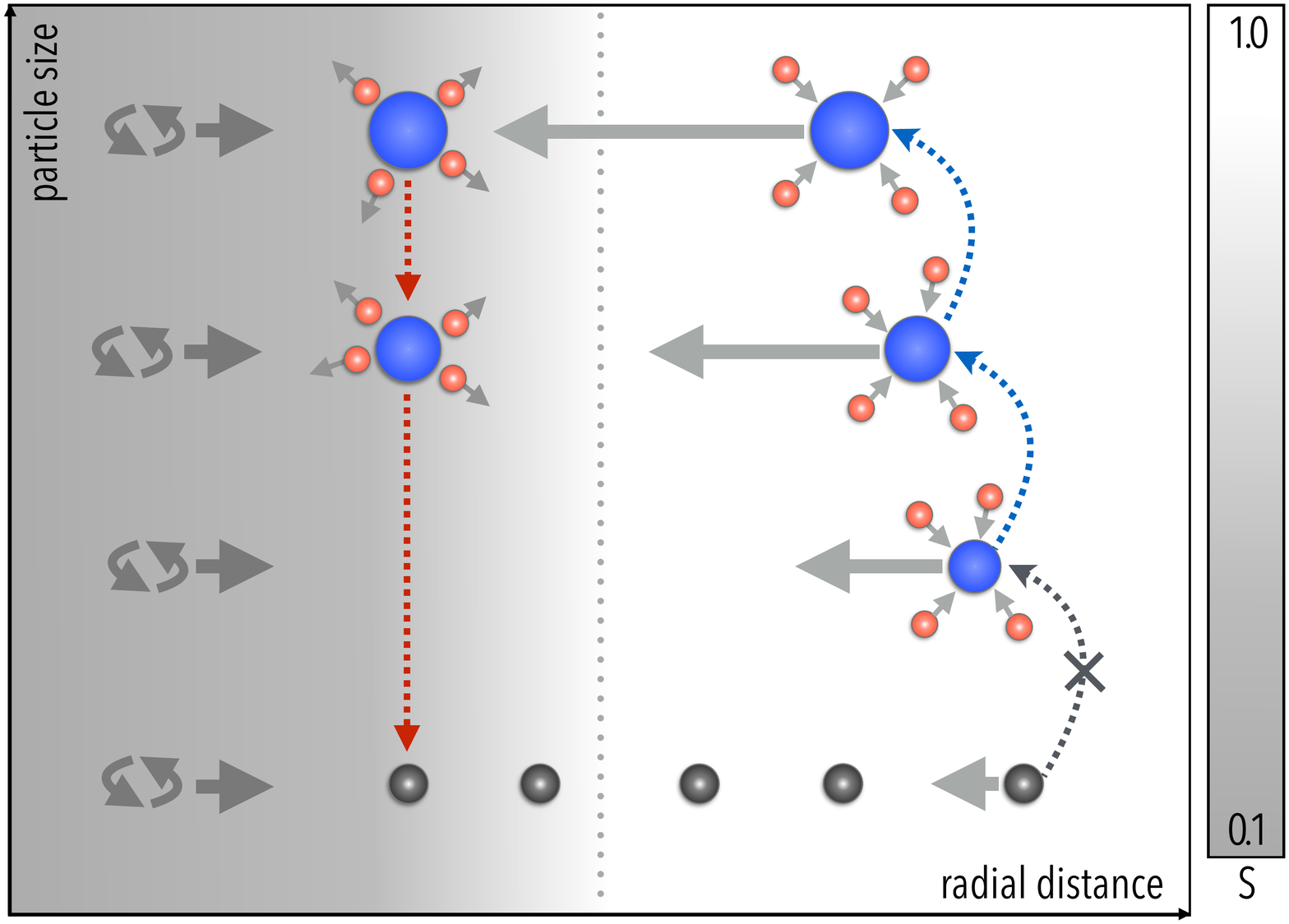}}
  \caption{Sketch of the physical processes included in our model. Ice-covered particles (blue) grow by the deposition of vapour (red) in a region where the saturation ratio is unity. Ice particles crossing into a region where the saturation ratio is lower than unity sublimate and, if they stay in this region long enough, eventually leave behind their silicate cores (grey) in addition to the sublimated vapour. Bare dust grains cannot acquire a new ice mantle since the saturation ratio does not reach the critical saturation ratio required for heterogenous nucleation. Dynamical processes are represented by grey arrows, with the left-directed arrows representing the size-dependent radial drift, and the arrows pointing to the right representing the outwards-directed part of the turbulent diffusion. Collisions are not included in this model.} 
              \label{fig:sketch}%
    \end{figure}
	
\begin{figure*}
	\centering
   \includegraphics[width=18cm]{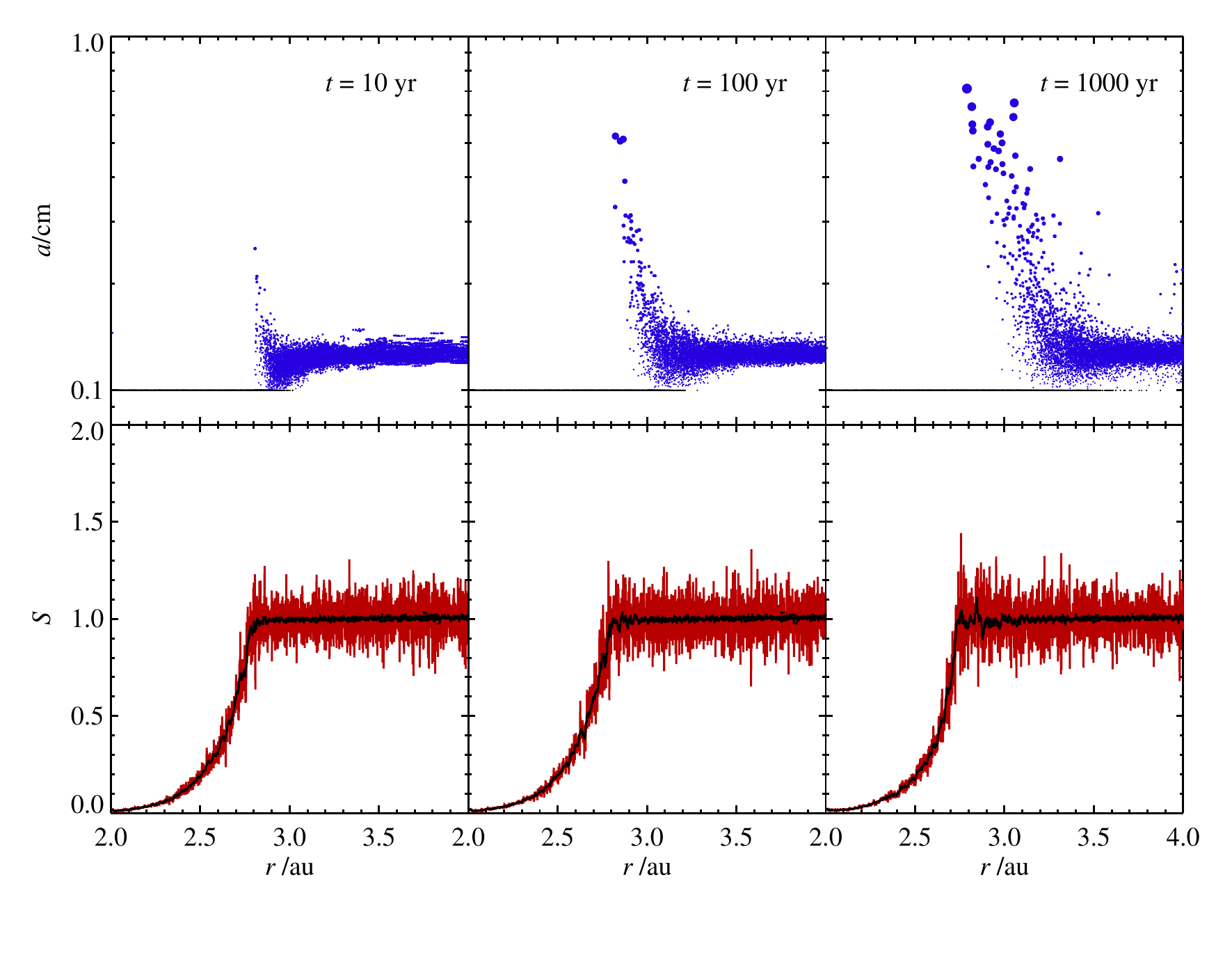}
     \caption{Particle growth for the fiducial full temperature-dependent nucleation model, where we include the influence of both depositional growth and heterogeneous nucleation on silicate dust. {\it Upper panel:} particle sizes as a function of semi-major axis at $t=10$, 100, and 1000 years. Ice-covered particles are shown as blue filled circles, with sizes proportional to the actual particle sizes, and the bare millimetre-sized silicate particles are shown in black. {\it Lower panel:} saturation ratio at corresponding times, where the black line is a smoothed average of the saturation ratio in each grid cell, shown in red. Growth takes place in a narrow region outside the ice line at $r_{\rm ice}\approx2.6\,{\rm au}$, with particles reaching centimetre sizes in 1000 years. Particle growth at larger $r$ is the result of vapour deposition at the start of the simulation.} 
     \label{fig:fp_dual}
\end{figure*}

\subsection{Heterogeneous nucleation, depositional ice growth, and sublimation}
\label{sec:nucleationcondensationandsublimation}
    
In this section we discuss the concepts of heterogeneous ice nucleation, or the build-up of the first monolayer of ice on bare silicate particles, and continued deposition on already ice-covered particles. A detailed description of how this is modelled is given in Appendix \ref{app:condensation}. 

We are interested in the behaviour of water around the ice line, the distance from the star where the total pressure of all locally available water molecules equals the saturated vapour pressure, and its influence on particle growth in this region. The position of the ice line is obtained naturally by comparing the partial pressure of water $P_{\rm v}$ to the saturated vapour pressure $P_{\rm sat}$, with the saturation ratio defined as
\begin{equation}
S=\frac{P_{\rm v}}{P_{\rm sat}}\,.
\end{equation}
Vapour pressure is a function of the amount and temperature of the available vapour and can be expressed using the ideal gas law as
\begin{equation}
P_{\rm v}=\frac{{\rm k_B} T}{m_{\rm v}} \rho_{\rm v}
\end{equation}
where $\rm k_B$ is the Boltzmann constant, $m_{\rm v}$ is the mass of a water molecule, and $\rho_{\rm v}$ is the density of vapour. The saturated vapour pressure is given by the Clausius-Clapeyron equation, and yields
\begin{equation} 
P_{\rm sat}=6.034\times10^{12} \,{\rm g \, cm^{-1} \,s^{-2}} \times {\rm e}^{-{5938 {\rm K}}/{\rm T}}
\end{equation} 
using experimentally determined material constants for water \citep{haynesetal1992}. Higher saturated vapour pressures are expected over nano-crystalline ice, crystallised from amorphous ice at low temperatures (below $T\approx160\,{\rm }K$), however the exact value of $S$ does not change the qualitative outcome of our model \citep{nachbaretal2018}. For a system with water vapour and water ice, $S=1$ means that the vapour is saturated, which is equivalent to having an equilibrium between the two phases. For $S>1$ any available vapour will be deposited and for $S<1$ sublimation will take place to restore $S=1$, if there is any ice available to sublimate. In Fig.\ \ref{fig:sr} we show the saturated vapour pressure as a function of semi-major axis in our simulation region, with the position of the ice line $r_{\rm ice}\approx2.6\,{\rm au}$, where $S=1$, marked. At $t=0$, there is an excess of vapour in the colder part of the region that gives a saturation ratio steeply increasing with semi-major axis. This vapour is quickly deposited on available surfaces and a saturation ratio of 1 is maintained outside the ice line, as is shown in the data for $t=10\,{\rm yr}$. 
    
Above, we have outlined a simple deposition and sublimation scenario, where we ignore the difference between heterogeneous nucleation, or the build-up of the first ice layer on a silicate particle, and continued depositional growth, when an ice mantle is already formed. In this simple model, we assume water vapour to be deposited on any available surface when $S>1$,  since the critical saturation ratio for deposition $S_{\rm crit}$ is equal for ice-covered and bare silicate particles with $S_{\rm crit, \, ice}=S_{\rm crit, \, sil}=1$. Thus, in any grid cell where $S>1$, water vapour is rapidly deposited to give thin ice mantles on as many of the available particles as possible, without differentiating between bare and icy particles, until equilibrium is reached.

In the more realistic scenario considered here, we take into account that it is energetically favourable for vapour to be deposited on an already ice-covered surface, as compared to building up the first monolayer of ice on a silicate particle. In our full nucleation model we do this by implementing experimental results of deposition on different surfaces, with $S_{\rm crit, \,ice}=1$ and $S_{\rm crit, \, sil}$ being temperature dependent. \citet{iracietal2010} found 
\begin{equation}
S_{\rm crit, \, sil}=-0.0626 \, T+13.0\,
\label{eq:scritsilfullTrange}
\end{equation} 
for the temperature range around the ice line, $160\,{\rm K}\leq{\rm T}\leq179\,{\rm K}$. As the exact form of the curve outside of this temperature range does not impact the outcome of our model, we extrapolate this result to lower and higher $T$, without allowing the critical saturation ratio to decrease below unity. Thus, we use
\begin{equation}
S_{\rm crit, \, sil}=
\begin{cases}
-0.0626 \, T+13.0 & T < 208\,{\rm K}\\	
1 & T \ge 208\,{\rm K}
\end{cases}
\label{eq:scritsil}
\end{equation} 
to cover the full range of temperatures in our simulations. We use data for silicon, which has similar deposition properties to silicate dust present in protoplanetary discs. In our model, heterogenous nucleation always results in an ice mantle covering the whole particle and, similarly, continued depositional growth results in a symmetric growth of the icy mantle so that particles are always spherical. We have also fitted data for Arizona Test Dust, a silicate material, from \citet{iracietal2010} using CNT with a distribution of contact angles \citep{wheelerbertram2012}, to yield a relationship between the critical saturation ratio and temperature in Appendix \ref{app:nucleation}. In Fig.\ \ref{fig:srt} we show the critical saturation ratios obtained by \citet{iracietal2010} for ice-covered and bare silicate particles as a function of semi-major axis and temperature throughout the simulation region. The water ice line, where $S=1$, is also shown. Since $S_{\rm crit, \,sil}>S_{\rm crit, \,ice}$ everywhere outside the ice line, the supersaturation needed for heterogeneous nucleation is always at least twice that of continued deposition. Comparing Figs.\ \ref{fig:sr} and \ref{fig:srt} we see that the critical saturation ratio for heterogeneous nucleation is higher than the actual saturation ratio everywhere outside the ice line at $t>0$. After the initial settling of vapour we therefore expect deposition on ice-covered particles, but no significant heterogeneous nucleation on bare silicate grains. 

   \begin{figure}
    \resizebox{\hsize}{!}{\includegraphics{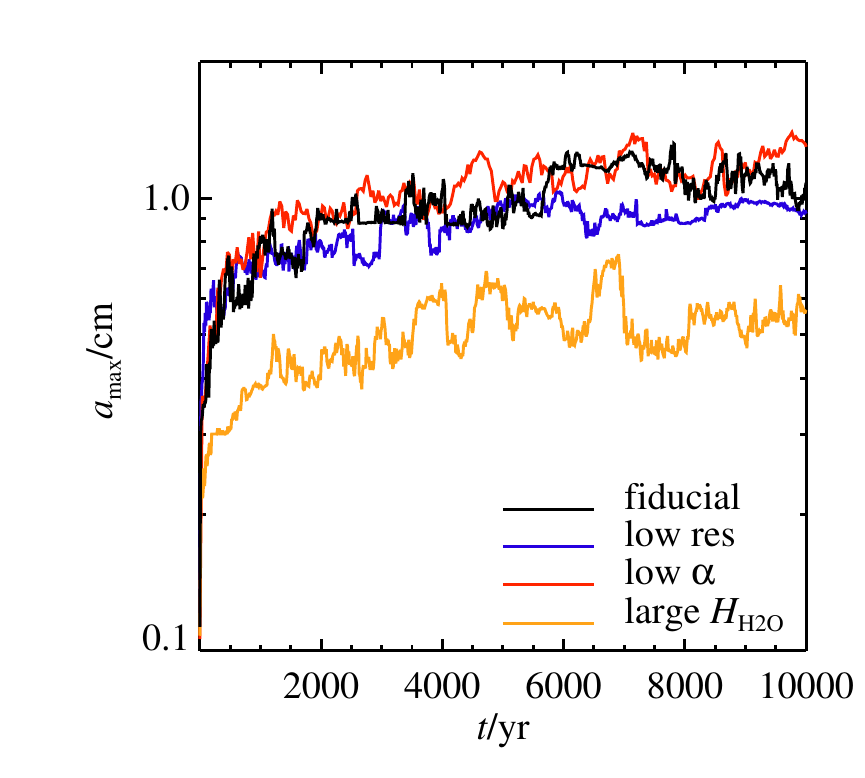}}
   \caption{Particle growth as a function of time in the full temperature-dependent nucleation model, with the nominal run shown as a black line. The largest particles reach centimetre sizes within a few 1000 years, and stay centimetre sized throughout the simulation. For comparison we also show a low-resolution run ($n_{\rm p}=10^5$) in blue, and a run with low turbulence ($\alpha=10^{-4}$) in red, both of which are in agreement with the nominal run. A run with a larger vapour scale height, $H_{\rm H_2O}=H_{\rm g}$, is shown in yellow.}
              \label{fig:ts_hr_alpha}%
    \end{figure}
    
 \begin{figure}
    \resizebox{\hsize}{!}{\includegraphics{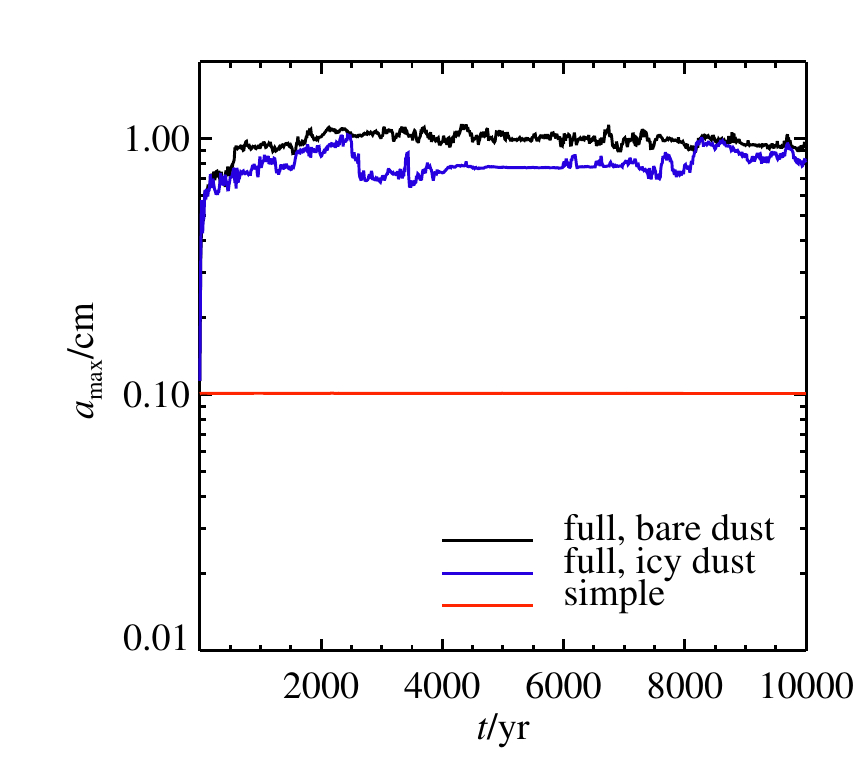}}
   \caption{Comparison of particle growth in the full temperature-dependent nucleation model and in the simple model, when including both micrometre-sized silicate grains and millimetre-sized ice-covered pebbles. In the simple model, shown in red, we ignore the difference between the critical saturation ratios needed for heterogeneous nucleation and vapour deposition. Vapour is then deposited on the micrometre-sized dust grains where most of the surface area resides, and the millimetre-sized particles do not grow. In the full nucleation model, denoted by a black line for initially bare dust and a blue line for initially ice-covered dust, heterogeneous ice nucleation of vapour on bare silicate particles is not possible, as the supersaturation required for heterogeneous nucleation is never reached. As a result, the icy pebbles grow quickly from millimetre to centimetre sizes.}
   \label{fig:ts_single_dual}
    \end{figure}
    
 \begin{figure*}
\resizebox{\hsize}{!}{\includegraphics{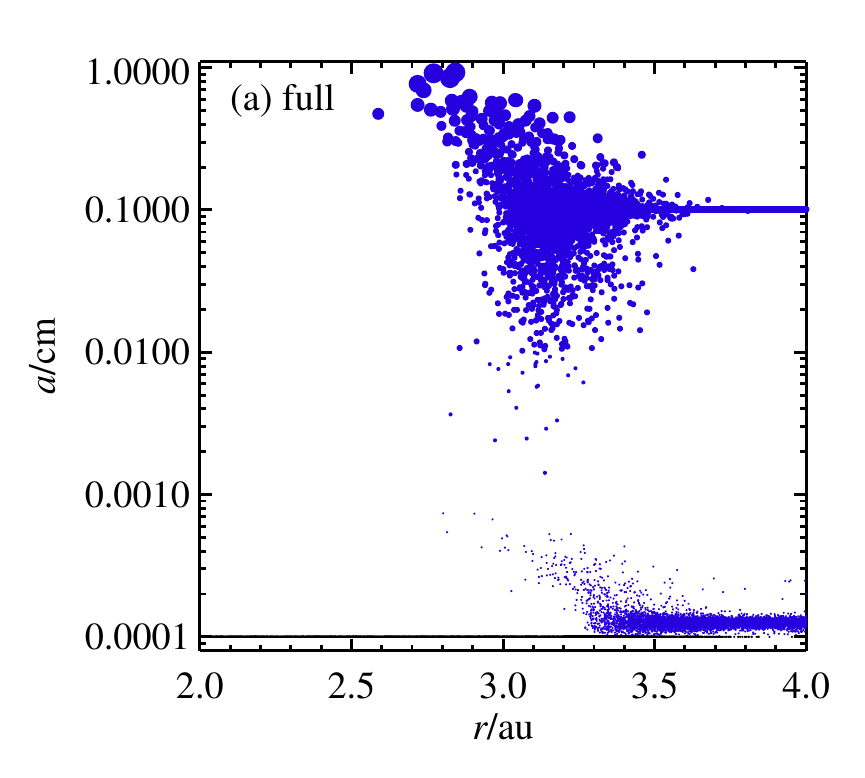}\includegraphics{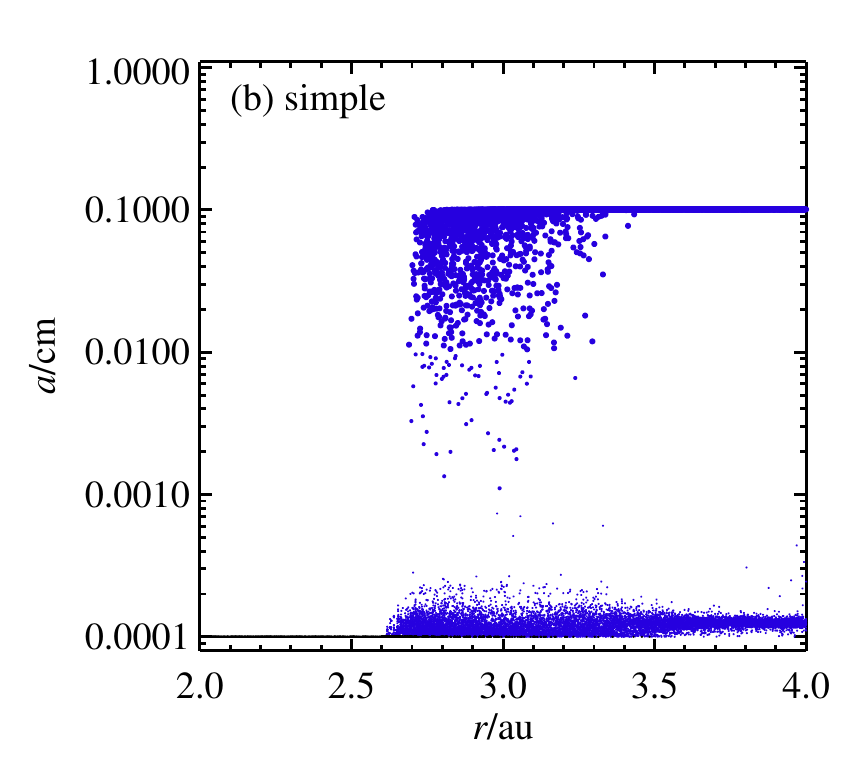}}
   \caption{Particle sizes at $t=1000\,\rm yr$ when starting from micrometre-sized and millimetre-sized icy particles. The micrometre-sized bare silicate grains are shown in black, and ice-covered particles in blue. {\it Panel a:} results from the full temperature-dependent nucleation model, in which the critical saturation ratio for heterogeneous nucleation is higher than for vapour deposition. Although a large amount of dust is present, the suppression of heterogeneous nucleation allows for growth only of the ice-covered particles, which quickly reach centimetre sizes. {\it Panel b:} results from the simple model, in which we ignore that a higher critical saturation ratio is needed for heterogenous nucleation than for vapour deposition. For micrometre-sized particles, growth by vapour deposition dominates, whereas for millimetre-sized pebbles the dominating process is sublimation. Thus, the available vapour is shared between the large amount of dust grains present, and larger particles do not grow.} 
   \label{fig:fp_dust}
    \end{figure*}
	
 \begin{figure*}
    \resizebox{\hsize}{!}{\includegraphics{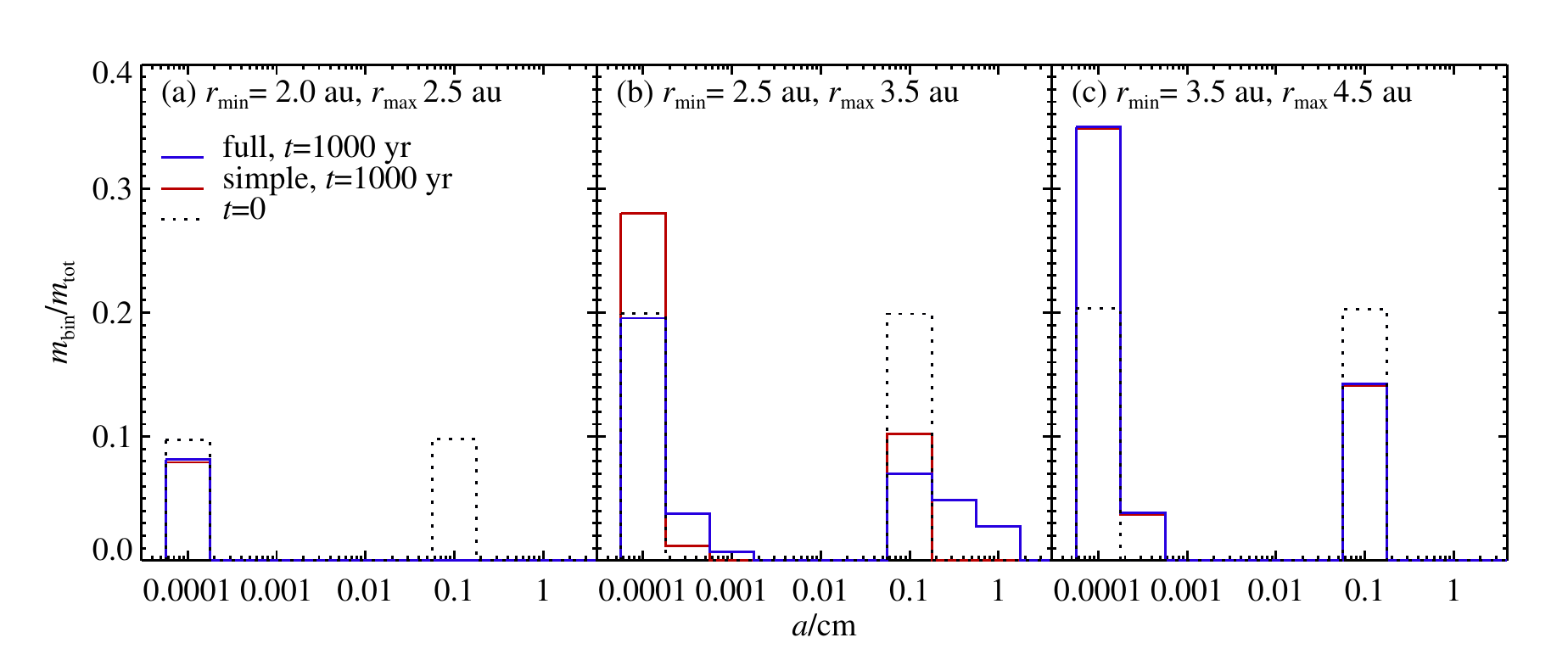}}
   \caption{Mass-weighted particle size distribution for the full temperature-dependent nucleation model (blue) and the simple model (red), for three radial sections of the simulation domain. A black dotted line shows the initial state, where the mass is divided equally between micrometre-sized particles with a thin ice mantle and millimetre-sized particles consisting of a micrometre-sized silicate nucleus and a thick ice mantle. The mass in solids in each radius bin  and in the full simulation region is denoted by $m_{\rm bin}$ and $m_{\rm tot}$, respectively. ({\it a}) In the region interior to the ice line, sublimation dominates and thus all solids at $t=1000\,{\rm yr}$ are in the form of bare, micrometre-sized dust grains, regardless of the model. ({\it b}) Particle growth in the region around the ice line is model dependent: In the full model both dust and pebbles grow, with a significant fraction reaching centimetre sizes, whereas in the simple model mainly micrometre-sized particles grow, reaching sizes of at most a few microns. ({\it c}) In the outer region, particle growth is inefficient for both models and the size distribution here is instead set by inwards drift of millimetre-sized particles and turbulent diffusion of micrometre-sized dust.}
   \label{fig:partdist}%
    \end{figure*}  
	

\section{Results}
\label{sec:results}
       
We ran and compared the results from our full temperature-dependent nucleation model and our simple model, with all the parameters that were varied shown in Table \ref{table:1}. Figure \ref{fig:sketch} gives an overview of the physical processes included in our model. 

\subsection{Particle growth including heterogeneous ice nucleation and depositional growth}
In Fig.\ \ref{fig:fp_dual} we show the state of the radial distributions of sizes and saturation ratio of the fiducial full nucleation model at 10, 100, and 1000 years. We started the simulation with millimetre-sized silicate cores covered by icy mantles randomly distributed across the simulation domain. 

As the simulation evolves, sublimation of the icy mantles provides the region inside of the ice line with vapour that can diffuse outwards and be redeposited on solid particles. Three regions can be identified. In the innermost part, where $P_{\rm v}<P_{\rm sat}$, no net deposition takes place and the icy mantles effectively sublimate as particles drift inwards. Outside of this region the vapour pressure is high enough for vapour to be deposited on ice, but not yet high enough for heterogeneous nucleation to take place, with $P_{\rm sat}<P_{\rm v}<P_{\rm nuc}$. Therefore we can find both bare silicates and quickly growing ice particles in this region. The bare silicates diffuse outwards as time progresses, whereas the large ice particles are more prone to inwards drift. In the outermost part of the simulation domain, $P_{\rm sat}<P_{\rm v}<P_{\rm nuc}$ just as in the second region. However, due to the efficiency of the deposition process, most vapour is deposited on ice particles already in the second region, leaving only a small amount of vapour to be deposited on the particles in the outer part, resulting in a negligible particle growth here.  

Vapour diffusing outwards across the ice line is rapidly deposited on ice
particles. Some of these icy particles subsequently cross back across the ice line and
sublimate there, while others diffuse outwards towards the colder regions of the disc. The radial
distribution of ice particles nevertheless does not spread as a Gaussian, due to
the destructive boundary condition at the ice line. 

The corresponding maximum particle size as a function of time is shown in Fig.\ \ref{fig:ts_hr_alpha}. Particle growth continues efficiently up to a centimetre, where the system settles in an equilibrium between growth and inwards drift and sublimation across the ice line. For comparison, we also include a low-resolution run with $n_{\rm p}=2\times10^5$ particles, and a run with $\alpha=10^{-4}$, corresponding to a less turbulent system, both of which are consistent with the fiducial model.

Throughout this paper we set the scale height for dust and vapour to $0.1H_{\rm g}$, assuming that the deposition timescale is shorter than the timescale to diffuse out over a full gas scale height, giving a vertical water distribution outside of the ice line that is shaped by the sedimenting pebbles. For comparison, we have included a run in  Fig.\ \ref{fig:ts_hr_alpha} where we instead assume instantaneous redistribution of water vapour and dust over the entire gas scale height at sublimation, setting $H_{\rm H2O}=H_{\rm g}$. Even in this case, icy pebbles grow, however the resulting maximum particle sizes are reduced by a factor of two compared to the nominal case.

\subsection{Comparing the simple model and the full nucleation model}
\label{sec:comp}

In Fig.\ \ref{fig:ts_single_dual} we compare particle growth for the simple and full temperature-dependent nucleation model. We started these simulations with two populations of particles, millimetre-sized pebbles consisting of a micron-sized silicate core and a thick ice mantle, and micrometre-sized dust, where the grains are covered by thin ice mantles. This is to highlight the effect that dust in discs has on inhibiting growth to larger sizes. For the full nucleation model we ran two versions, one where all dust grains are covered by thin ice mantles ($1\%$ of the radius of a silicate grain) already at the start of the simulation, and one where dust particles are introduced as bare silicate grains. However, in the latter case the excess vapour gives a vapour pressure in the beginning of the simulation that is high enough also for nucleation on bare silicates everywhere where deposition is possible, except for in the region $2.6\, \rm au\lesssim r \lesssim 2.7\,\rm au$. Thus the two cases give similar results, with both mimicking an inflow of icy dust from the outer disc. In the simple scenario, the maximum particle size does not increase from the initial $a_{\rm init}=1\, {\rm mm}$, since most of the surface area is made up of small dust grains and consequently all the vapour is deposited on the dust, leading to a very small growth of individual particles. In the full nucleation model, however, heterogeneous nucleation on dust particles requires a saturation ratio significantly above unity, as can be seen in Fig.\ \ref{fig:srt}, whereas depositional growth of ice-covered particles takes place as soon as $S>1$. As is clear from Figs.\ \ref{fig:sr} and \ref{fig:fp_dual}, the saturation ratio never reaches much above unity, which means that vapour mostly is deposited on icy surfaces instead of nucleating on bare silicate dust. The silicate dust that is left after sublimation of icy pebbles crossing the ice line will thus remain bare, whereas deposition takes place only on the remaining icy particles. This results in an efficient growth, not affected by the abundance of dust grains present, with ice particles reaching centimetre sizes in 1000 years, similar to in Fig.\ \ref{fig:ts_hr_alpha}. 

A snapshot at $t=1000\rm \,yr$, comparing the full nucleation model and the simple model, is shown in Fig.\ \ref{fig:fp_dust}, where a) is the full nucleation model and b) is the simple model. In both models, the particle population just below a millimetre is due to partial sublimation of ice mantles of the initially millimetre-sized particles, whereas the particle populations consisting of particles just above a micrometre and, for the full model, above a millimetre, are due to growth by vapour deposition and nucleation. 

We show the mass-weighted particle size distribution in the whole simulation domain, divided into three radial sections, in Fig.\ \ref{fig:partdist}. In the inner and outer regions, both models behave similarly. Inside the ice line, sublimation of ice mantles gives a dominating population of bare dust grains, and in the outer part of the simulation region, particle growth by nucleation and deposition is inefficient since vapour released at the ice line is deposited locally. The particle population in the outer region is therefore set mainly by dynamics: millimetre-sized particles tend to drift inwards, whereas turbulent diffusion allows micrometre-sized dust grains to move outwards. In the region around the ice line at $r_{\rm ice}\approx2.6\,\rm au$, the models behave distinctly different with the inhibition of nucleation in the full model allowing growth of icy particles to proceed to large sizes, whereas in the simple model vapour is mainly deposited on the large surface area represented by micrometre-sized dust grains.


\section{Discussion}
\label{sec:discussion}

\subsection{Model assumptions}

In this paper we have focused on investigating particle growth due to deposition and sublimation around the water ice line in a dynamical model. We have included the important layering of ice particles, with a silicate nucleus and an icy mantle, and taken into account the difference between heterogeneous nucleation, or the build-up of the first monolayer of ice, and continued depositional growth on already ice-covered particles in our full temperature-dependent nucleation model. In order to isolate these effects, our model has a number of simplifications and assumptions that we will discuss in this section.

\paragraph{\it Disc evolution} We model a protoplanetary disc in a fixed state, and do not take the time evolution of the disc and ice line into account. As the timescale of growth by deposition is very short, $\tau_{\rm dep}\approx1000\, {\rm yr}$, compared to the lifetime of the disc of a few million years, we find this a reasonable assumption. The ice line can also move outwards during time periods as short as a few years due to stellar outbursts such as FU Orionis events \citep{ciezaetal2016}, leaving a region of supersaturated vapour that can be redeposited and build up large icy pebbles, as suggested by \citet{hubbard2017}. In a future work we will investigate such a scenario in more detail.

\paragraph{\it Initial particle size} We start our simulations with millimetre-sized pebbles, and in some cases micrometre-sized dust, with the size of the pebbles motivated by experiments on particle coagulation \citep{blumwurm2008, guttleretal2010}. However, this maximum size can be seen as a conservative choice, as drift-limited growth is suggested to yield centimetre-sized aggregates for a disc model comparable to ours \citep{birnstieletal2012}. We found that increasing the initial maximum particle size in our model does not change the relative particle growth, with the largest resulting particles having radii approximately one order of magnitude larger than the initial maximum particle size.

\paragraph{\it Particle collisions} We do not include collisions between particles in our model, which means that particle sizes are set only by deposition and sublimation after the initial state of the simulation. Typically, laboratory experiments and simulations of coagulation find maximum particle sizes of a millimetre, which is consistent with the size of ice nuclei in our simulations. The number and outcome of collisions, however, is important not only for setting the maximum particle size available, but also for determining the distribution of particle sizes arising from fragmenting collisions. The particle size distribution, in turn, affects the deposition growth outcome. Depending on the relative distribution of solid surface area between small and large particles, deposition can result in thin ice mantles spread over many small dust particles in the case of a dominating fraction of small particles, or thicker mantles and growth to larger sizes in the case of most surface area being provided by larger particles. The size distribution outcome of particle collisions in an equilibrium between coagulation and fragmentation can be described by a power law giving the number density distribution as
\begin{equation}
\frac{{\rm d}n}{{\rm d}a} \propto a^{-q}\,,
\end{equation}
where $n$ is the number of particles and $a$ is the particle radius. Here $q=3$ gives equal surface in equal logarithmic size intervals, which implies that deposition takes place on all grain sizes, whereas $q>3$ leads to a scenario where most vapour is deposited on the small fragments. The steady state solutions of \citet{birnstieletal2011} yield various outcomes in terms of the size distribution of the solids; under some assumptions the surface area is dominated by small grains and under others by large grains. Our results are valid both in cases where fragmentation can be ignored and when it leads to a top-heavy size distribution. We note that our ice particles grown by vapour deposition will take the form of solid crystalline ice and that such monolithic particles should have very different fragmentation properties from the aggregates considered in \citet{birnstieletal2011}.

As we do not account for collisions, we also ignore sintering, the effect by which dust grains can fuse together at temperatures just below sublimation \citep[e.g.][]{sirono1999}. This process can reduce the sticking efficiency of dust and has therefore also been shown to lead to disc structures with observable dark and bright rings in the vicinity of ice lines \citep{okuzumietal2016}.

\paragraph{\it Particle structure} We assume spherical particles that consist of a silicate nucleus and acquire an icy mantle when heterogeneous nucleation of water vapour takes place. A more complex model including for example porosity, dust aggregation, and different ice nuclei compositions is beyond the scope of this paper, as is the inclusion of active sites, local features on the particle surface where ice growth starts \citep{kiselevetal2017}. Although all of these factors likely are of importance for heterogeneous ice nucleation and ice growth, we would expect such effects to be important only for a small fraction of the particles and hence that this will not change the qualitative outcome of our model.

\paragraph{\it Kelvin curvature effect} The spatial separation of ice-covered pebbles and bare silicate dust in the disc that we find in this paper is due to the higher saturated vapour pressure required for the onset of heterogeneous nucleation on silicate. This effect would be even more pronounced if the dependence of deposition on particle size were taken into account. The Kelvin curvature effect is the increase in saturated vapour pressure due to the curvature of the substrate, and yields
\begin{equation}
  P_{{\rm s},i} = P_{{\rm s},0} \exp(a_{\rm K}/a_i), 
\end{equation}
with the Kelvin radius 
\begin{equation}
  a_{\rm K} = \frac{2 \gamma_{\rm S} V_{\rm m}}{\mathcal{R} T} \, ,
\end{equation}
where $\gamma_{\rm S}$ is the ice-vapour surface tension, $V_{\rm m}$ is the molar volume, and $\mathcal{R}$ is the universal gas constant. 
This is approximately 0.6 nm for water ice, and hence the saturated vapour pressure
over a micron-sized grain is 0.1\% higher than over a flat surface. The
growth rate by deposition and sublimation is
\begin{equation}
  \dot{m} = 4\pi a^2 \,v_{\rm th}\, \rho_{\rm v}\left( 1-\frac{P_{\rm sat}}{P_{\rm v}} \right)\,,
\end{equation}
where $v_{\rm th}$ is the thermal velocity of vapour. Assuming spherical particles and writing $S=P_{\rm v}/P_{\rm sat}$ we can rewrite the growth rate in terms of radius as
\begin{equation}
  \dot{a} =v_{\rm th}\, \frac{\rho_{\rm v}}{\rho_\bullet} (1-S^{-1})\,.
\end{equation}
At $S=0.999$ felt by micron-sized grains, the rate at which the ice mantle is lost by sublimation is $10^{-12}$ m/s at the ice line, where $\rho_{\rm v}=10^{-8}\,{\rm kg\,m^{-3}}$.
Micron-sized grains near the water ice line thus lose their ice mantles after only $10^6$ s, or a few days, while grains of 0.1 mm can retain their ice mantles for $10^8$ s, that is, several years. Small ice grains are therefore not efficient at `stealing' water vapour, since they quickly lose their ice mantles even if $S$ is high enough for heterogeneous ice nucleation to take place.

\subsection{Implications and future work}
We find that solids are divided into two different populations at the ice line. Just outside the ice line the mass is dominated by fast-growing icy pebbles that reach centimetre sizes on a timescale of 1000 years, whereas particles stay small and silicate-heavy further out from the ice line. The growth of icy pebbles in narrow regions around ice lines could be consistent with some of the observations of ringed structures in several protoplanetary discs \citep{almapartnershipetal2015, andrewsetal2016, isellaetal2016, tsukagoshietal2016, ciezaetal2017,vanboekeletal2017, andrewsetal2018}. In the disc around HL Tau, emission dips coinciding with major ice lines were found \citep{almapartnershipetal2015}, and interpreted by \citet{zhangetal2015} as the result of fast pebble growth due to deposition of volatiles. The innermost dip in the disc of HL Tau at $r\approx13\,{\rm au}$ corresponds to the water ice line at $r\approx2.6\,{\rm au}$ in our model, with the temperature profiles differing due to the high accretion rate of HL Tau, estimated to $\dot{M}\approx10^{-7}\,M_\odot \, {\rm yr^{-1}}$ \citep{becketal2010}. 

In several discs such emission dips have been found to match the location of the CO ice line at  around $30\, {\rm au}$ with $T_{\rm cond}\approx 20\,\rm K$ \citep{zhangetal2016}. Although in this paper we focus on the water ice line at $r_{\rm ice}\approx2.6\,{\rm au}$, we expect similar behaviour at other ice lines. Our model can be extended to include other volatiles, with the same principles applying further out in the disc, where the $\rm {CO}_2$ ice line at about $10\, {\rm au}$ with $T_{\rm cond}\approx 70\,\rm K$ is the next major one outside of the water ice line. Just as for the water ice line, we here need to distinguish between heterogeneous nucleation and further depositional growth, since experimental evidence shows that a higher saturation ratio is required for $\rm {CO}_2$ vapour to nucleate on water ice than for it to be deposited on particles already covered by $\rm {CO}_2$-ice \citep{glandorfetal2002, nachbaretal2016}. Applying our model to ice lines further out in the disc will be the subject of future investigations.

A few emission dips have been suggested to be a result of dynamical clearing by growing planets \citep{isellaetal2016, ciezaetal2017}. This would be a natural second step of the fast pebble growth at ice lines that we report here, in agreement with earlier studies showing that growth to planetesimals is facilitated at ice lines \citep{rosjohansen2013, schoonenbergormel2017, drazkowskaalibert2017}.  

The separation of growing icy particles and silicate-dominated dust could also explain the low water content in comets that has been found by analysing cometary interiors \citep{kuppersetal2005}. The large icy pebbles in our model do not readily diffuse outwards as the turbulent diffusion is counteracted by radial drift inwards. However, the small dust grains are not as affected by drift and can therefore diffuse outwards in the disc. This could result in a large-scale separation of water and silicates, where the water content of the disc increases close to the ice line and decreases in the outer parts of the disc, leaving less water available in the cold regions where comets are formed. We will explore this separation of ice and silicates further in future global simulations run over a longer timescale. 
%

\section{Conclusions}
\label{sec:conclusions}

In this paper we have investigated heterogeneous ice nucleation and vapour deposition at the water ice line, and included two important effects previously often ignored when modelling dust growth in protoplanetary discs: the release of dust grains upon sublimation of icy particles and the distinction between heterogeneous nucleation of the first ice layer on a silicate particle and continued depositional growth of already ice-covered particles. 

Small dust grains can in principle function as ice nuclei, competing with the growing icy pebbles for the available water vapour, resulting in a small net growth by deposition. This scenario is demonstrated in our simple model, where we do not make the important distinction between heterogeneous nucleation and further depositional ice growth. In such a simplified model the competition between the abundant dust and ice grains results in negligible particle growth. 

However, experiments on heterogeneous ice nucleation, performed to understand ice cloud formation in the Martian atmosphere and applicable to protoplanetary disc conditions, have shown that a substantially higher water vapour pressure is needed for the first ice layer to form, as compared to continued deposition of water vapour on ice \citep{iracietal2010}. Implementing these experimental results leads to a scenario where icy particles can efficiently grow to pebble-sizes, whereas silicate dust particles stay small and without ice mantles. This is shown in our full temperature-dependent nucleation model, where we find that solids are divided into two populations at the ice line. Just outside the ice line the mass is dominated by icy pebbles, growing to centimetre sizes on a timescale of 1000 years, whereas micron-sized silicate dust makes up the largest part of the mass further out in the disc. 

Our model highlights growth by vapour deposition and we have ignored changes to particle sizes by other means, most importantly coagulation and fragmentation. A fundamental assumption in this model is that the large pebbles that grow outside the ice line do not create a large population of fragments that would dominate the total surface area. We argue that these pebbles are monolithic condensates and that such particles in contrast to aggregates would not readily form small fragments in collisions. However, verifying these results in a complete dust coagulation model, including coagulation, fragmentation, and the distinction between heterogeneous ice nucleation and vapour deposition is an important future extension of this work.

\begin{acknowledgements}
The authors would like to thank the referee, Joanna Dr\c{a}\.zkowska, for her  comments that helped us improve the paper. Helene Stalheim and Hanna Vehkam{\"a}ki are acknowledged for discussions on CNT and ice nucleation. KR and AJ acknowledge funding by the European Research Council (ERC Starting Grant 278675-PEBBLE2PLANET). AJ was also supported by the Knut and Alice Wallenberg Foundation (Wallenberg Academy Fellow grant 2012.0150 and grant 2014.0048 ”Bottlenecks for particle growth in turbulent aerosols”) and the Swedish Research Council (grant 2014-5775). IR and DS were supported by AtmoRemove.
\end{acknowledgements}

 \bibliographystyle{aa} 
 \bibliography{refs} 
 
\begin{appendix}

\section{Turbulent diffusion}
\label{app:diffusion}

\begin{figure*}
  \centering
  \includegraphics[width=0.8\linewidth]{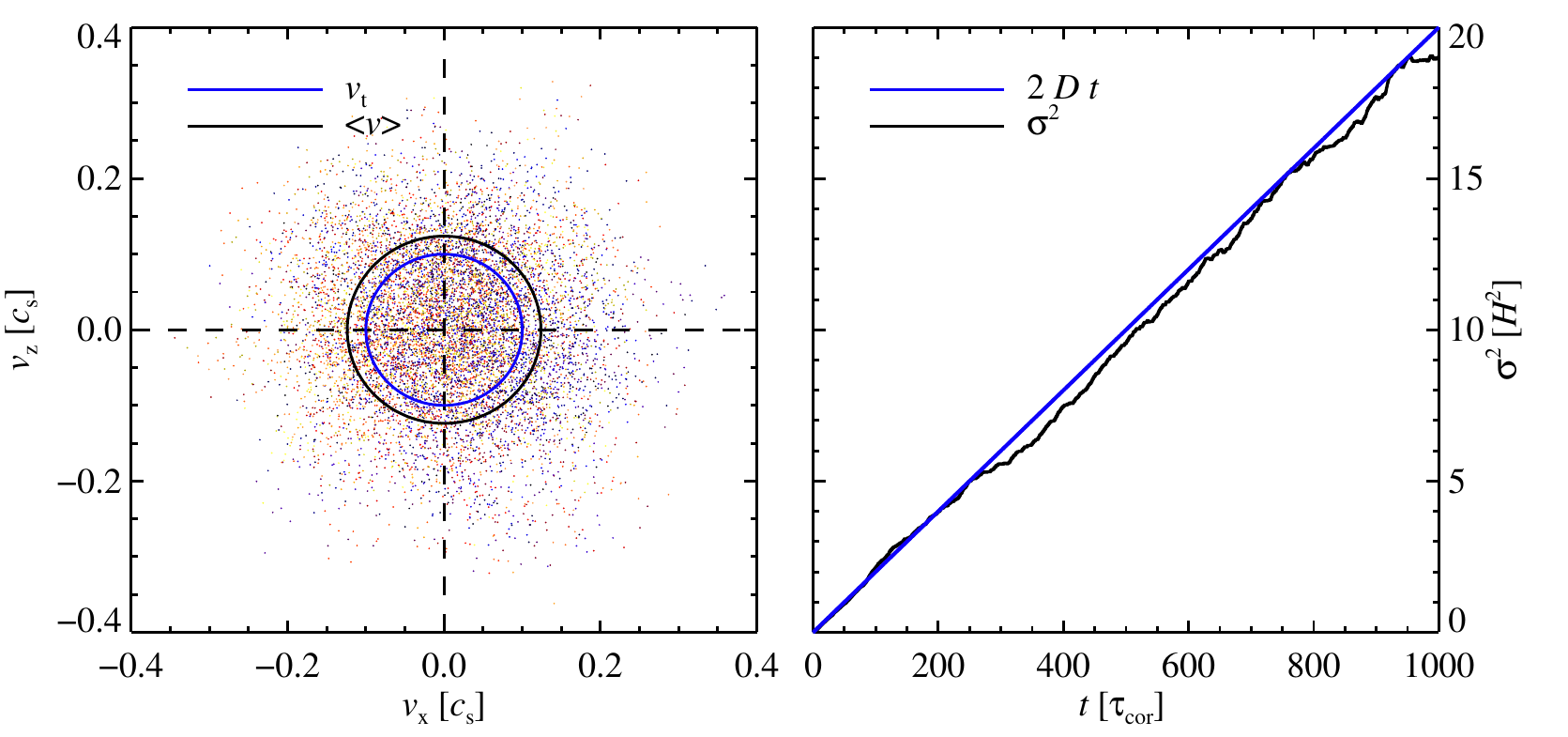}
  \caption{Velocity components and squared width of particle positions in the damped random walk model. {\it Left panel:} radial and vertical velocity components
  resulting from the damped random walk model. The turbulent speed $v_{\rm
  t}=\sqrt{\alpha} c_{\rm s}$ is indicated with a blue circle and the mean
  measured speed with a black circle. {\it Right panel:} square of the
  standard deviation of the $x$-position of 1000 particles undergoing a damped
  random walk (black line). We compare to the squared ensemble width from the
  desired diffusion coefficient $D=\alpha c_{\rm s}^2 \varOmega^{-1}$.}
  \label{fig:rwd}
\end{figure*}

The turbulent diffusion of particles accelerated by drag from the turbulent gas
can be mimicked by a random walk process. In \cite{rosjohansen2013} we let the
particles move with the turbulent speed $v_{\rm t} = \sqrt{\alpha} c_{\rm s}$,
in the direction $\theta$ relative to the radial direction, over the correlation
time $\tau_{\rm cor}=\varOmega^{-1}$.  However, such an approach implies that
the particles move
in straight trajectories over several grid cells before their direction of
travel is changed instantaneously. In this paper we instead achieve the desired
value of the turbulent diffusion coefficient by forcing the acceleration of the
particles on the timescale $\tau_{\rm for}$ and damping the turbulent particle
velocity on the correlation timescale $\tau_{\rm cor}$. The acceleration of the
turbulent velocity field $(v_x,v_z)$ is given by
\begin{eqnarray}
  \dot{v}_x &=& f \cos{\theta} - \frac{1}{\tau_{\rm cor}} v_x \, , \\
  \dot{v}_z &=& f \sin{\theta} - \frac{1}{\tau_{\rm cor}} v_z \, .
\end{eqnarray}
Here $x$ denotes the radial direction and $z$ the vertical direction in the protoplanetary disc. This equation system describes a damped random walk, with forcing acceleration
$f$ and with $\theta$ chosen randomly between $0$ and $2 \pi$ every forcing time
$\tau_{\rm for}$. The velocity field will achieve statistical equilibrium at the
characteristic value \citep[see discussion in][]{johansenetal2009}
\begin{equation}
  v_{\rm t} \sim \sqrt{\tau_{\rm for} \tau_{\rm cor}} f \, .
\end{equation}
The correlation time of the velocity field, that is, the timescale over which a
particle maintains a coherent direction, will be approximately $\tau_{\rm
cor}$. This will result in a turbulent diffusion coefficient of the particles of
\begin{equation}
  D = \frac{v_{\rm t}^2}{\tau_{\rm cor}} \, .
\end{equation}
This expression only contains the correlation timescale $\tau_{\rm cor}$, not
the forcing timescale $\tau_{\rm for}$. Therefore we can choose any forcing
timescale with the appropriate forcing 
\begin{equation}
  f= \frac{2 v_{\rm t}}{\sqrt{\tau_{\rm for} \tau_{\rm cor}}} \, .
\end{equation}
We multiply here the forcing by a factor of two, in order to obtain an equilibrium
value of $v_{\rm t}$ in better agreement with the desired value. We choose the
forcing timescale to be equal to the time-step of the code in order to have
smooth particle trajectories and to avoid the additional bookkeeping involved
in choosing longer time-steps. Setting then $v_{\rm t} = \sqrt{\alpha} c_{\rm
s}$ and $\tau_{\rm cor}=\varOmega^{-1}$ yields the desired diffusion
coefficient
\begin{equation}
  D = \alpha c_{\rm s}^2 \varOmega^{-1} \, .
\end{equation}
We show in Fig.\ \ref{fig:rwd} the components of the turbulent velocity arising
from the damped random walk model as well as the squared width, $\sigma^2$, of the positions
of 1000 particles undergoing a damped random walk. The squared width follows the
diffusion coefficient through the expression
\begin{equation}
  \sigma^2 = 2 D t \, .
\end{equation}
We find excellent agreement between the measurement and the analytical
expression.

The diffusion operator on a passive scalar $\rho_{\rm p}$ appears as
\begin{equation}
  \dot{\rho}_{\rm p} = \vc{\nabla} \cdot \left[ D \rho_{\rm g}
  \vc{\nabla} \left( \frac{\rho_{\rm p}}{\rho_{\rm g}} \right) \right] \, ,
\end{equation}
where $\rho_{\rm p}$ and $\rho_{\rm g}$ are the particle and gas density, respectively. 
Writing out in two terms for constant $D$ yields
\begin{equation}
  \dot{\rho}_{\rm p} = D \nabla^2 \rho_{\rm p} - \vc{\nabla} \cdot \left( D
  \rho_{\rm p} \vc{\nabla} \ln \rho_{\rm g} \right) \, .
  \label{eq:rhopdiff2}
\end{equation}
The first term is mimicked by the damped random walk described above. However,
that term in isolation would yield constant $\rho_{\rm p}$ in equilibrium. In
reality turbulent diffusion strives towards constant $\rho_{\rm p}/\rho_{\rm
g}$; this is ensured by the second term in Eq.\ (\ref{eq:rhopdiff2}). We
therefore add an additional speed to the particles, $\vc{v}_{\rm diff}=D
\vc{\nabla} \ln \rho_{\rm g}$. The gradient of the gas density can be calculated
directly from the gas density structure
\begin{equation}
  \rho_{\rm g} = \frac{\varSigma_{\rm g}}{\sqrt{2\pi}H} \exp[-z^2/(2 H^2)] \, .
\end{equation}
Setting $\varSigma_{\rm g}$ and $c_{\rm s}=H \varOmega$ as power laws with
index $q_\varSigma$ and $q_c$ yields the logarithmic gradient
\begin{eqnarray}
  \frac{\partial \ln \rho_{\rm g}}{\partial r} &=& \frac{1}{r} \left[
  \left( \frac{z}{H} \right)^2 (q_c+1.5) + q_\varSigma - (q_c + 1.5) \right] \,
  , \\
  \frac{\partial \ln \rho_{\rm g}}{\partial z} &=& - \frac{z}{H^2} \, .
\end{eqnarray}

\section{Scheme for sublimation and deposition}
\label{app:condensation}

We describe here our algorithm to solve the growth equation for the mass $m_i$
of particles in swarm $i$,
\begin{equation}
  \dot{m}_i = 4 \pi a_i^2 v_\perp \left( \rho_{\rm v} - \rho_{\rm s}
  \right) \, .
\end{equation}
Here $a_i$ is the particle radius, $v_\perp$ is the average speed of water
vapour flowing through a surface, $\rho_{\rm v}$ is the vapour mass density, and
$\rho_{\rm s}$ is the saturated vapour density of water at the environmental
temperature. The water vapour speed is
\begin{equation}
  v_\perp = \sqrt{\frac{k_{\rm B} T}{2 \pi m_{\rm v}}} \, .
\end{equation}
In terms of radius, this gives
\begin{equation}
  \dot{a}_i = \frac{v_\perp}{\rho_\bullet} \left( \rho_{\rm v} - \rho_{\rm s}
  \right) \, .
\end{equation}
Here $\rho_\bullet$ is the material density and we assumed spherical particles.
Mass conservation implies that the total density $\rho_{\rm tot} = \rho_{\rm v}
+ \rho_{\rm p}$ (where $\rho_{\rm p}$ is the mass density of solid particles) is
conserved in sublimation and deposition processes. Therefore we can rewrite the
growth equation as
\begin{equation}
  \dot{a}_i = \frac{v_\perp}{\rho_\bullet} \left( \rho_{\rm tot} - \rho_{\rm
  p} - \rho_{\rm s} \right) \, .
\end{equation}
Now we write $\rho_{\rm p}$ in terms of $a_i$ in order to close the system. That
yields
\begin{equation}
  \dot{a}_i = \frac{v_\perp}{\rho_\bullet} \left( \rho_{\rm tot} - \sum_j (4
  \pi/3) \rho_\bullet a_j^3 n_j - \rho_{\rm s} \right) \, .
\end{equation}
Here $n_i$ is the number density of particles represented by swarm $i$. This
simple-looking equation actually represents $N$ mutually coupled non-linear
differential equations. However, we note that the right-hand side is shared
between all the equations, since the index $j$ only appears under the summation.
Hence we can write $a_i(t) = a_{0i} + \Delta a(t)$, where $\Delta a(t)$ is
independent of $i$. The growth equation for $\Delta a$ becomes
\begin{equation}
  \dot{\Delta a} = \frac{v_\perp}{\rho_\bullet} \left( \rho_{\rm tot} -
  \sum_j (4 \pi/3) \rho_\bullet (a_{0j} + \Delta a)^3 n_j - \rho_{\rm s} \right)
  \, .
\end{equation}
This equation can be solved by separation of variables,
\begin{equation}
  \frac{{\rm d} \Delta a}{f(\Delta a)} = {\rm d} t \, .
\end{equation}
Here $f(\Delta a)$ is a third order polynomial in $\Delta a$ whose coefficients
are easily calculated. We expand $f(\Delta a)$ in terms of its three roots
$\alpha_1$, $\alpha_2$, $\alpha_3$ as
\begin{equation}
  f = C_3 (\Delta a - \alpha_1) (\Delta a-\alpha_2) (\Delta a - \alpha_3) \, .
\end{equation}
Here $C_3$ is the coefficient in front of the $(\Delta a)^3$ term. We
furthermore expand the inverse polynomial as
\begin{equation}
  \frac{1}{f} = \frac{1}{C_3} \left( \frac{\beta_1}{\Delta a - \alpha_1} +
  \frac{\beta_2}{\Delta a - \alpha_2} + \frac{\beta_3}{\Delta a - \alpha_3}
  \right) \, .
\end{equation}
The coefficients $\beta_1$, $\beta_2$, and $\beta_3$ are obtained by matching the
coefficients of the polynomial and solving a linear equation system. Now we may
integrate the inverse growth equation (left-hand side from 0 to $\Delta a$ and
right-hand side from 0 to $\Delta t$) to obtain
\begin{equation}
  \frac{1}{C_3} \left[ \beta_1 \ln \left(\frac{\Delta a-\alpha_1}{-\alpha_1}
  \right) + \beta_2 \ln \left(\frac{\Delta a-\alpha_2}{-\alpha_2} \right) +
  \beta_3 \ln \left(\frac{\Delta a-\alpha_3}{-\alpha_3} \right) \right] = \Delta
  t \, .
\end{equation}
This equation now connects $\Delta t$ and $\Delta a$ algebraically.
Unfortunately the equation is transcendental in $\Delta a(\Delta t)$, but it can
be solved numerically by bisection. This way we can solve for
deposition and sublimation locally in a grid cell along the analytical curve and
avoid time-stepping. The situation is nevertheless made more complicated by the
presence of a silicate core in each solid particle. We take the core radius
$a_{{\rm c}}$ to be constant for all particles species, generally 1 mm. However,
at sublimation we are not allowed to integrate past the core radius. In that
case we integrate until the first grain is bare and then exclude this grain from
the subsequent sublimation integration and so on with the following grains.

\section{Heterogeneous ice nucleation and classical nucleation theory}
\label{app:nucleation}
Experimental data by \citet{iracietal2010} for the critical saturation ratio $S_\mathrm{crit}$ as a function of the nucleation temperature $T_\mathrm{nuc}$ used in the simulations are here fit within the framework of classical nucleation theory \citep[CNT,][]{pruppacheretal1998,seinfeldpandis2012} . 

Under the simplifying assumption of heterogeneous ice nucleation taking place on a flat surface, we derive 
\begin{equation}
	S_\mathrm{crit}(T_\mathrm{nuc}; \alpha) = \exp{\left[\frac{f(\alpha)}{a\cdot b\cdot T^3_\mathrm{nuc}}\right]}\,,
	\label{eq:Scrit}
\end{equation}
with the geometric term $f(\alpha)=(2+\cos(\alpha))(1-\cos(\alpha))^2/4$ depending on the contact angle $\alpha$ and constants $a:=-k\,\ln\left(J/J_0\right)$ and $b:=\frac{3\,R^2\,\rho^2}{16\,\pi\,M^2\,\zeta}$, where $k=1.38064852\cdot 10^{-23}\,\mathrm{J}\,\mathrm{K}^{-1}$ is the Boltzmann constant, $J=1\,\mathrm{cm}^{-2}\,\mathrm{s}^{-1}$ the nucleation rate threshold chosen, $J_0=10^{16}\,\mathrm{cm}^{-2}\,\mathrm{s}^{-1}$; $R = 8.314\,\mathrm{J}\,\mathrm{mol}^{-1}\,\mathrm{K}^{-1}$  is the universal gas constant, $\rho=0.92\,\mathrm{g\,cm}^{-3}$ the mass density of ice, $M = 18.015\,\mathrm{g\,mol}^{-1}$ the molar mass of H$_2$O, and $\zeta=1.065\cdot 10^{-5}\,\mathrm{J\,cm}^{-2}$ the vapour-ice surface tension. 

\begin{figure}[h!]
	\centering
	\includegraphics[width=88mm]{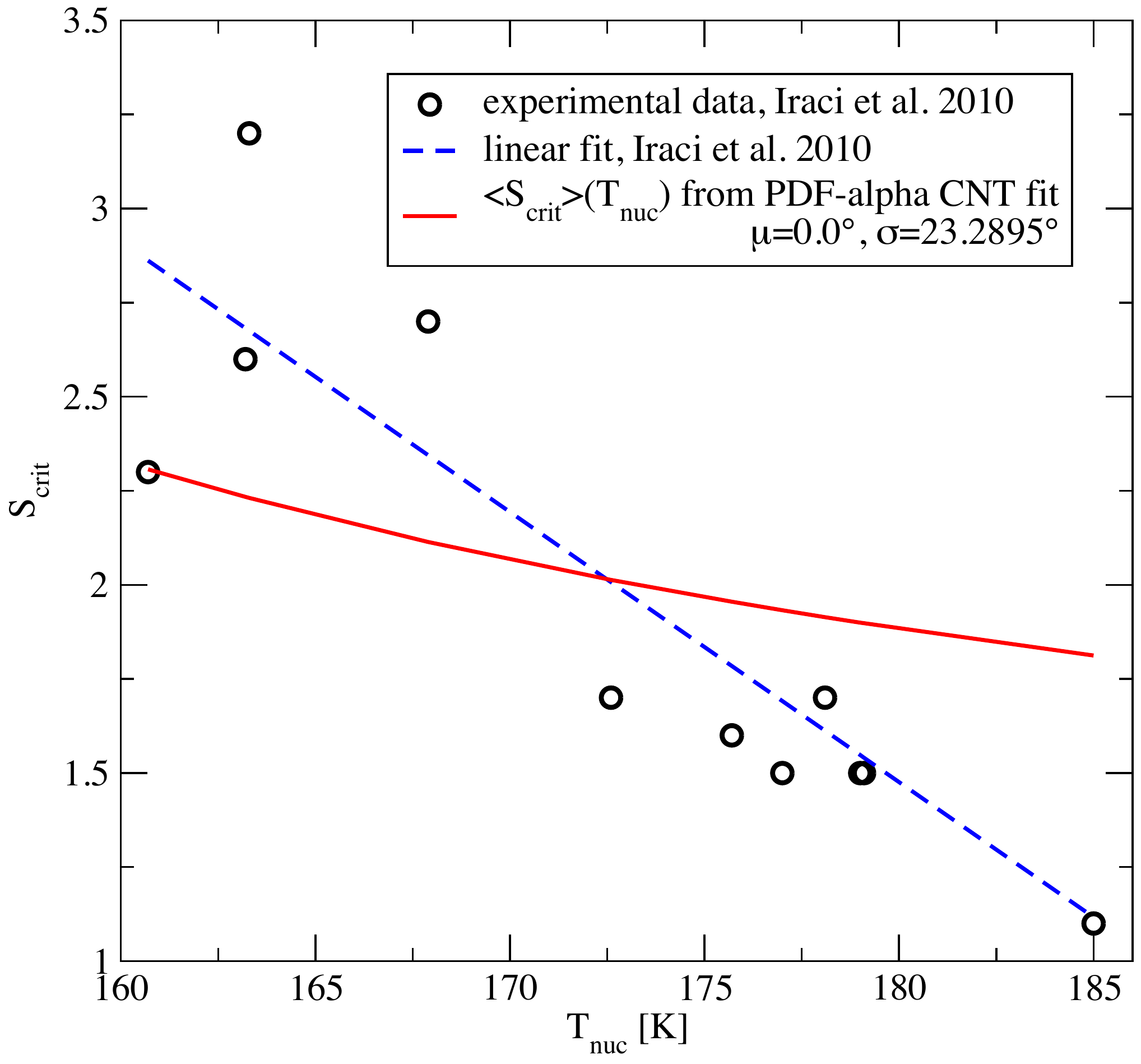}
	\caption{\label{fig:fittoiraci} Critical saturation ratio as a function of nucleation temperature for Arizona Test Dust \citep{iracietal2010}.}
\end{figure}

\citet{wheelerbertram2012} have shown that deposition nucleation on mineral dust particles can be described by extensions of CNT, among them a model employing a distribution of contact angles (PDF-$\alpha$ CNT) \citep{luondetal2010, marcollietal2007}, rather than by CNT with a single contact angle $\alpha$. Taking the PDF-$\alpha$ CNT approach, we assume a Gaussian distribution of contact angles $\phi(\alpha; \mu, \sigma)$ (normalised on $[0,\pi]$), calculate the expectation value 
\begin{equation}
	\langle S_\mathrm{crit}\rangle_\alpha (T_\mathrm{nuc}; \mu, \sigma)  = \int^\pi_0 S_\mathrm{crit}(T_\mathrm{nuc}; \alpha) \phi(\alpha; \mu, \sigma) d\alpha \,, 
	\label{eq:Scritexp}
\end{equation}
and use Eq.~\ref{eq:Scritexp} to fit the experimental data for Arizona Test Dust reported by \citet{iracietal2010} with mean $\mu$ and standard deviation $\sigma$ of the angle distribution function as free parameters. The result of the least-squares fit is shown in Fig.~\ref{fig:fittoiraci}.

The parameters obtained from the fit, $\mu=0.0^\circ$ and $\sigma=23.2895^\circ$, are found to favourably compare to those obtained by \citet{wheelerbertram2012} for kaolinite ($\mu=0.0^\circ$ and $\sigma=54.14^\circ$) and illite ($\mu=35.43^\circ$ and $\sigma=14.64^\circ$). 
 
We note that, although the PDF-$\alpha$ CNT reproduces the right order of magnitude of supersaturation ($s:=S-1\sim 100 \%$), the temperature dependence of the theoretical description is too weak when compared to the experimental data. This is due to the functional form derived from CNT and does not depend on the fitting parameters. Furthermore, deviations of this kind are to be expected as it is well known that CNT generally fails to capture the temperature dependence of the nucleation rates \citep{vehkamaki2006, vehkamakiriipinen2012}. While the theory detailed above describes the data qualitatively, we stress that CNT-based descriptions often yield unrealistic results due to oversimplifications in the theory's basic assumptions. Advanced theoretical descriptions are generally more reliable but more complex or computationally costly \citep{naparietal2010, oleniusetal2018}.

\end{appendix}

\end{document}